\documentclass[runningheads]{llncs}

\usepackage[utf8]{inputenc} 
\usepackage[T1]{fontenc}    
\usepackage{hyperref}       
\usepackage{url}            
\usepackage{booktabs}       
\usepackage{amsmath}
\usepackage{amsfonts}       
\usepackage{nicefrac}       
\usepackage{microtype}      
\usepackage{lipsum}		
\usepackage{graphicx}
\usepackage{doi}
\usepackage{subcaption}

\begin{document}

\title{Emergent kin selection of altruistic feeding via non-episodic neuroevolution}


\author{
    Max Taylor-Davies\inst{1} \and
    Gautier Hamon\inst{2} \and
    Timothé Boulet\inst{2} \and
    Clément Moulin-Frier\inst{2}
}
\authorrunning{M. Taylor-Davies et al.}

\institute{School of Informatics, University of Edinburgh, Edinburgh, Scotland \\
\email{m.taylor-davies@sms.ed.ac.uk}
\and Flowers team, Inria Center of the University of Bordeaux, France.
}

\maketitle

\date{}

\begin{abstract}
Kin selection theory has proven to be a popular and widely accepted account of how altruistic behaviour can evolve under natural selection. Hamilton's rule, first published in 1964, has since been experimentally validated across a range of different species and social behaviours. In contrast to this large body of work in natural populations, however, there has been relatively little study of kin selection \emph{in silico}. In the current work, we offer what is to our knowledge the first demonstration of kin selection emerging naturally within a population of agents undergoing continuous neuroevolution. Specifically, we find that zero-sum transfer of resources from parents to their infant offspring evolves through kin selection in environments where it is hard for offspring to survive alone. In an additional experiment, we show that kin selection in our simulations relies on a combination of kin recognition and population viscosity. We believe that our work may contribute to the understanding of kin selection in minimal evolutionary systems, without explicit notions of genes and fitness maximisation. 
\end{abstract}


\section{Introduction}\label{sec:intro}
At first glance, it seems difficult to square the phenomenon of purely altruistic behaviour (acts which confer a benefit to the recipient at a cost to the actor) with the basic principle of natural selection: how can a gene be selected for when it decreases, rather than increases, the fitness of its host? One plausible account can be made through the theory of inclusive fitness. Key to this theory is the recognition that individual organisms within a social environment are not isolated from their conspecifics in terms of fitness. Whether a given gene is selected for is thus determined by its effect(s) on the fitness of any bearers of copies of that gene. Under this view, we can think of an altruistic act as an exchange of fitness from one agent to another. If the exchange is positive-sum and both sides are bearers of the gene in question, then from the gene's perspective the behaviour confers a fitness benefit--even while it decreases the fitness of the acting individual. 

Hamilton's rule, published in 1964, offers a simple and elegant formalisation of this idea \cite{hamilton_genetical_1964_1,hamilton_genetical_1964_2}. Imagine a particular gene produces some social behaviour which confers fitness benefit $b$ to the recipient at a cost $c$ to the actor. Given a measure $r$ of the genetic relatedness between the two, the rule states that this gene should increase in frequency if $rb > c$. A key consequence of Hamilton's rule is that an altruistic behaviour that targets the actor's kin will undergo more positive selection than an equivalent behaviour which is indiscriminate or favours non-kin. This is referred to as `kin selection' \cite{smith_group_1964}, and can be seen as a more narrow form of inclusive fitness. Targeting of behaviour towards kin can be facilitated either by mechanisms of kin recognition, or more simply by population viscosity. Since the publication of Hamilton's rule, empirical studies across a range of different species \cite{bourke_hamiltons_2014} have confirmed its predictions for social behaviours such as guarding among female Allodapine bees and defence of dominant male Wild turkeys by non-dominant males \cite{stark_cooperative_1992,krakauer_kin_2005}. Kin selection theory has also been used to explain insect eusociality \cite{hamilton_genetical_1964_2,hughes_ancestral_2008} and allomothering in certain species of monkey \cite{fairbanks_reciprocal_1990}, and has even been invoked in accounts of language evolution \cite{fitch_kin_2004,fitch_evolving_2007}. 

In this paper, we demonstrate an emergent kin selection mechanism within a population of continuously evolving artificial agents situated in an ecologically plausible environment. We place thousands of agents within a large gridworld filled with resources that regenerate through time (Figure 1). Each agent is controlled by its own artificial neural network, mapping its local observation of the environment to its possible actions: moving in space, consuming a resource, feeding the agent next to it, or reproducing. We depart from the standard evolutionary computation paradigm: our simulations include no explicit fitness measure, instead relying on a 'minimal criterion' for individual agents' survival and reproduction, governed by a simple physiological model.  In addition, we do not reset the agent population or environment state at any point during simulation (a framework referred to as non-episodic neuroevolution \cite{hamon_eco-evolutionary_2023}). Importantly, the only kin-specific mechanism we introduce is the ability of agents to distinguish between adults, infants and their own offspring. The answer to our main research question is therefore far from trivial: in these simple, ecologically plausible conditions, will agents evolve the altruistic behaviour of feeding their own offspring? It is important to note that doing so will not favour their own survival and reproduction in any way. It will only favour the propagation of their own (mutated) genome, but without any explicit incentive to do so. To study this research question, we manipulate certain features of the environment, indirectly varying the fitness tradeoff $rb - c$. Our results show that altruistic feeding behaviour emerges as this differential increases. We also show that an increase in the prevalence of the altruistic feeding behaviour is associated with an increase in its selectivity towards agents' own offspring. To our knowledge, this is the first demonstration that kin selection naturally evolves in realistic ecological conditions where artificial agents continuously evolve in a non-episodic environment.

\section{Environment and agents}\label{sec:environment}
\subsection{Environment dynamics}\label{subsec:env-dynamics}
The environment we use is based on the non-episodic neuroevolution framework described in \cite{hamon_eco-evolutionary_2023}. Simulations are run in a 100x100 gridworld which contains both food resources (plants) and agents, with each tile containing at most one agent, up to a maximum population size of 2500. Resources are generated and regenerated via simple spontaneous growth: at the start of simulation, each tile has probability $p=0.1$ of containing a plant; at every subsequent timestep, each non-plant-containing tile has constant probability $p=0.003$ of spontaneously spawning one. The environment is initialised with a starting population of 2000 agents, all with random neural network weights. Agents' survival and reproduction is governed entirely by simple physiological rules, to enable minimal-criterion neuroevolution \cite{brant_minimal_2017}. Agents are born with an initial energy of 65. They lose 0.1 energy passively at each timestep, with a further loss of 1.0 for every non-idle action they perform. An agent dies when its energy falls below 0. Agents can increase their energy (up to a maximum of 200) by consuming plants, with each plant worth 20 units of energy. If an agent's energy is at or above a threshold of 85, and the population is less than the maximum of 2500, it can use the reproduce action to produce a single offspring (at an energy cost of 30). Offspring spawn into the nearest empty square and receive a copy of their parent's neural network weights randomly mutated by the addition of zero-mean Gaussian noise, i.e. $\theta_\text{offspring} := \theta_\text{parent} + \mathcal{N}(0, \sigma)$ (for all simulation runs reported in this paper we use $\sigma = 0.02$). Agents younger than 100 timesteps are considered to be 'infants'--they are unable to reproduce, and may also experience slightly different environment dynamics or action mechanics (see Section \ref{sec:experiments} for more detail). 

\subsection{Agents}\label{subsec:agents}
\begin{figure}[t]
    \centering
    \includegraphics[width=\linewidth]{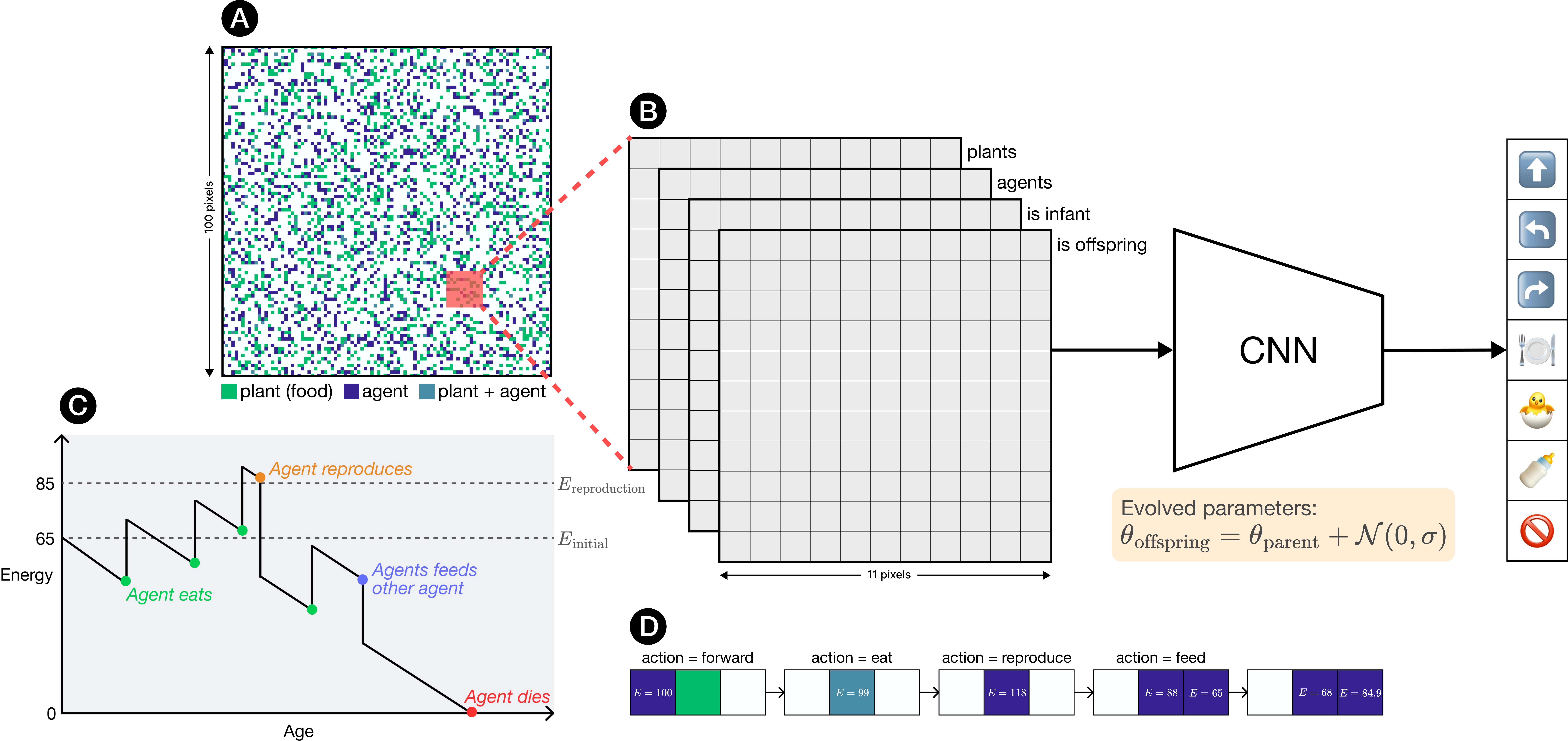}
    \caption{(A) A snapshot of the simulation environment. (B) an illustration of the agent architecture, observation and action spaces. Each agent receives an 11x11x4 observation of its local environment--this is fed through a convolutional neural network (CNN) to produce a vector of action probabilities. (C) an example illustration of the energy level over time for a single agent. (D) an illustration of a short action sequence in which an agent moves onto a food resource, eats the food resource, reproduces and then feeds their offspring.}
    \label{fig:env-agent-diagram}
\end{figure}

At each timestep, every agent in the population samples an action from a policy parameterised by an individual neural network (NN) controller, implemented as simple 3-layer convolutional neural network (CNN). The weights of each agent's controller remain constant throughout their lifetime--i.e. there is no learning, and behavioural adaptation can happen only through neuroevolution. Agents observe the environment within an 11x11 window centred at their current location, within which they see the locations of both plants and other agents. They can also see, for each other agent within this window, two binary values indicating whether that agent is A) an infant, and B) their own offspring. The action space is given by \{forward, turn left, turn right, eat, reproduce, feed, idle\}. The `eat' action has the effect of consuming a single plant resource, and is only effective if the agent is occupying a tile containing a food resource. The `feed' action transfers one food resource's worth of energy to a single other agent, and is only effective if the two agents are in adjacent tiles, and the feeder is directly oriented towards the feedee (facilitating some degree of selectivity in feeding behaviour). The `reproduce' action is as described above (Section \ref{subsec:env-dynamics}). Fig \ref{fig:env-agent-diagram} gives an illustration of the environment, agent architecture, and energy dynamics. 

\section{Experiments}\label{sec:experiments}
\begin{figure}
     \centering
     \begin{subfigure}[b]{0.49\textwidth}
         \centering
         \includegraphics[width=\textwidth]{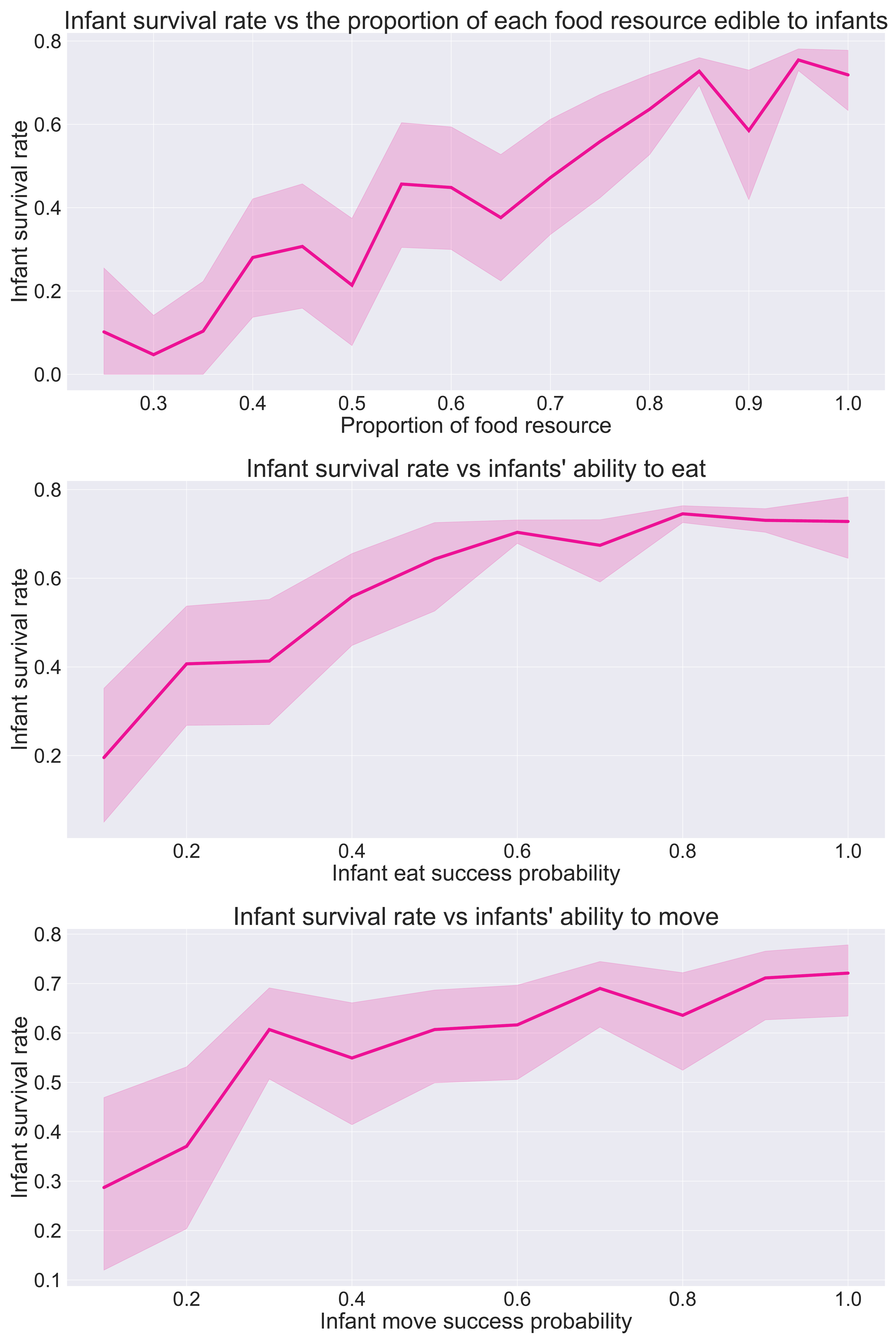}
         \caption{The proportion of agents that survive to adulthood, as a function of our three experimental parameters, measured over the final 10k timesteps of each 100k-timestep simulation, in environments where agents do \textbf{not} have the ability to feed one another.}
         \label{fig:infant-survival}
     \end{subfigure}
     \hfill
     \begin{subfigure}[b]{0.49\textwidth}
         \centering
         \includegraphics[width=\textwidth]{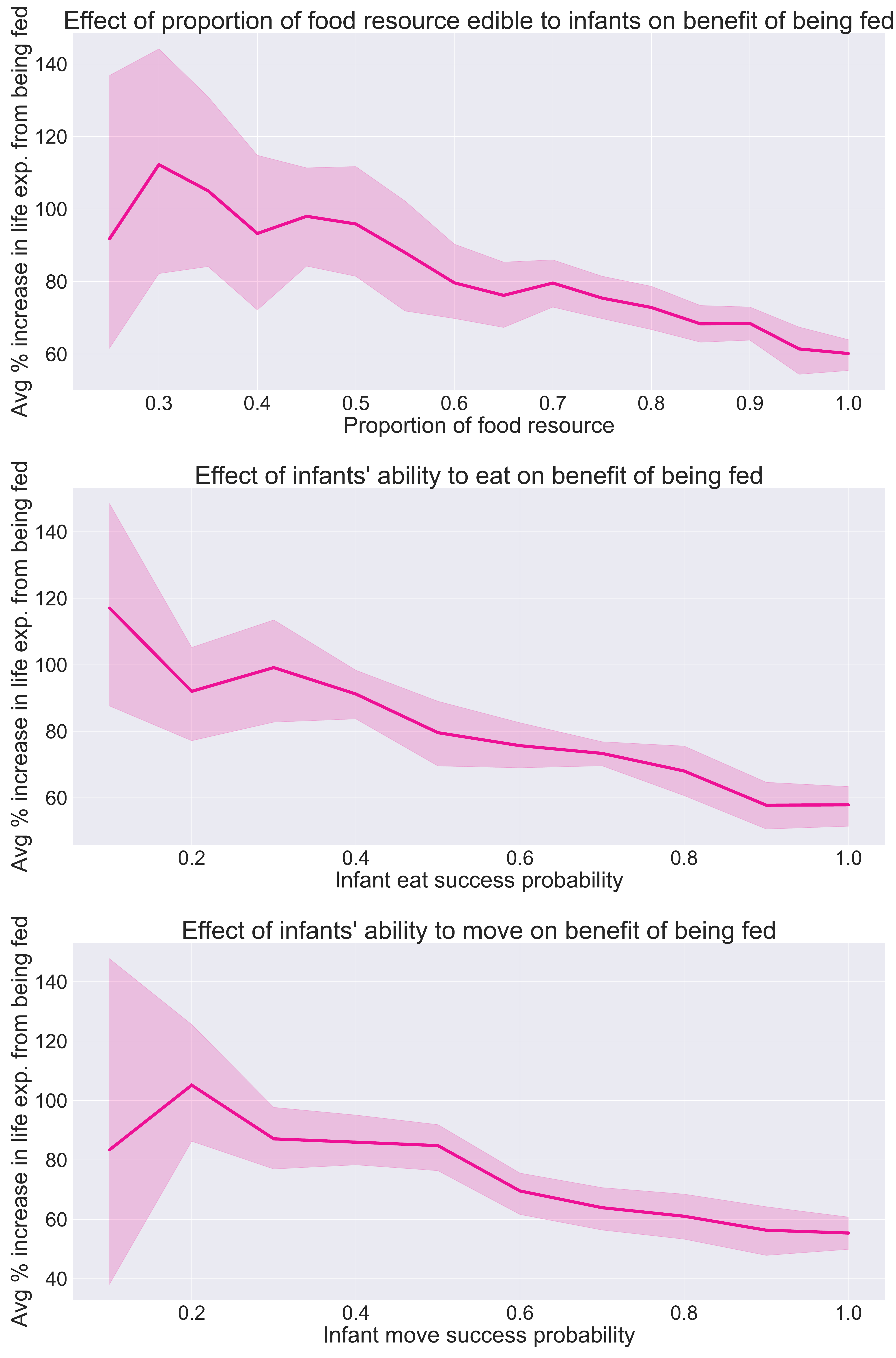}
         \vspace{-0.5mm}
         \caption{The estimated benefit (to life expectancy) of being fed as an infant, as a function of our three experimental parameters, measured over the final 50k timesteps of each 500k-timestep simulation}
         \label{fig:feeding-benefit}
     \end{subfigure}
     \caption{The effect of varying the three chosen simulation parameters on (a) the infant survival rate (ISR) with feeding disabled, and (b) the estimated benefit of being fed as an infant ($\hat{b}$), given by the average percentage increase in lifespan for infants that are fed relative to those that aren't. Error bars in all plots represent bootstrapped 95\% confidence intervals over 20 seeds per parameter value.}
     \label{fig:three graphs}
\end{figure}

In our experiments, we seek to determine whether altruistic feeding behaviour (i.e. the zero-sum transfer of resources from one agent to another) can emerge in our neuroevolution framework as a result of kin selection. We focus on a narrow version of kin selection that considers only the relationship between parents and their direct offspring. Since in this case the genetic relatedness $r$ is close to 1 (specifically, $1 - \sigma$), Hamilton's rule \cite{hamilton_genetical_1964_1} tells us that in our environment, resource transfer from parent to child should be selected for when the fitness benefit $b$ to the child of gaining the resource exceeds the fitness cost $c$ to the parent of losing the resource. Our hypothesis is thus that in simulation environments where $b - c$ is greater, we should observe both more feeding behaviour \emph{and} feeding behaviour that is more selectively directed towards agents' own offspring. If on the other hand there is no kin selection process, 

While controlling this tradeoff directly is challenging, we can take an indirect approach by varying the relative effect of a marginal food resource on the continued survival of parent and child. We do this by manipulating certain environment conditions that make it harder for \emph{infant} agents to survive to adulthood (and thus to reproductive age), while keeping the survival difficulty for adult agents unchanged. This should have the effect of increasing $b$ while keeping $c$ constant, and thus increasing the differential $b - c$. The parameters we manipulate for this are:
\begin{itemize}
    \item \textbf{infant food energy proportion}: the proportion of available energy that infant agents gain from each food resource (where adults get 100\%), defaulting to 1.0. For example, if this value is 0.5, then an infant's energy level will increase by 10 upon consuming a plant, whereas an adult's would increase by 20. 
    \item \textbf{infant eat success probability}: the probability with which an infant's `eat' action succeeds, assuming they are located at a food resource (where adults' `eat' actions are deterministically successful), defaulting to 1.0. For example, if this value is 0.5, an infant would have to use the `eat' action twice as often as an adult to get the same expected energy gain. 
    \item \textbf{infant move success probability}: equivalent to \textbf{infant eat success probability} but for the `move' action, defaulting to 1.0. For example, if this value is 0.5, an infant would move around the environment twice as slowly as an adult following the same action policy. 
\end{itemize}

To establish the effect of each of these parameters on the environment survivability, we first run a set of simulations with \textbf{feeding behaviour disabled} (i.e. agents' action spaces do not contain the `feed` action), and track the `infant survival rate` (ISR) over time. For a window $[t_1, t_2]$, we compute this as 
\begin{equation}
    \text{ISR} = 1 - \frac{\text{num infant deaths in } [t_1, t_2]}{\text{total num deaths in } [t_1, t_2]}
\end{equation}
Figure \ref{fig:infant-survival} shows the ISR recorded over the final 10k steps of 100k-step simulations as a function of each of the three parameters listed above, with 20 seeds run per parameter value (and the other parameters held constant at their default value of $1.0$). For these preliminary experiments, we chose to end simulations at 100k timesteps after observing minimal changes in the overall action distribution past this point (indicating that agents' policies had evolved to a stable equilibrium). 
While the trends are not monotonic, and there is a fair amount of variance over the 20 seeds, we can see that the recorded ISR increases with all of the three chosen parameters--as we would expect. 

Having established that our chosen parameters do affect infants' ability to survive to adulthood (and thus in theory the excess fitness $b - c$), we re-enable agents' ability to feed others, and run a new set of simulations over 20 seeds x 500k timesteps per parameter value. 

\section{Results}

\subsection{The benefit of being fed is inversely correlated with infants' ability to survive}
While we use the three simulation parameters listed above as experimental proxies for $b - c$, we would like to obtain a more direct estimate $\hat{b}$ of the fitness benefit to agents of being fed. Since agents that live longer will in expectation produce more offspring, we can do this by computing, over the final 50k steps of each simulation run, the average difference in lifespan between agents that do and do not receive feeding as infants. Fig \ref{fig:feeding-benefit} shows how $\hat{b}$ changes as a function of each of the three experimental parameters. Across all three parameters, we see that the benefit of being fed during infancy on agents' life expectancy (and thus ability to reproduce) is higher in environments that are more hostile to infants' survival. 

\subsection{Agents engage in more feeding and more selective feeding when the benefit of being fed is higher}
\begin{figure}[t]
    \centering
    \includegraphics[width=\linewidth]{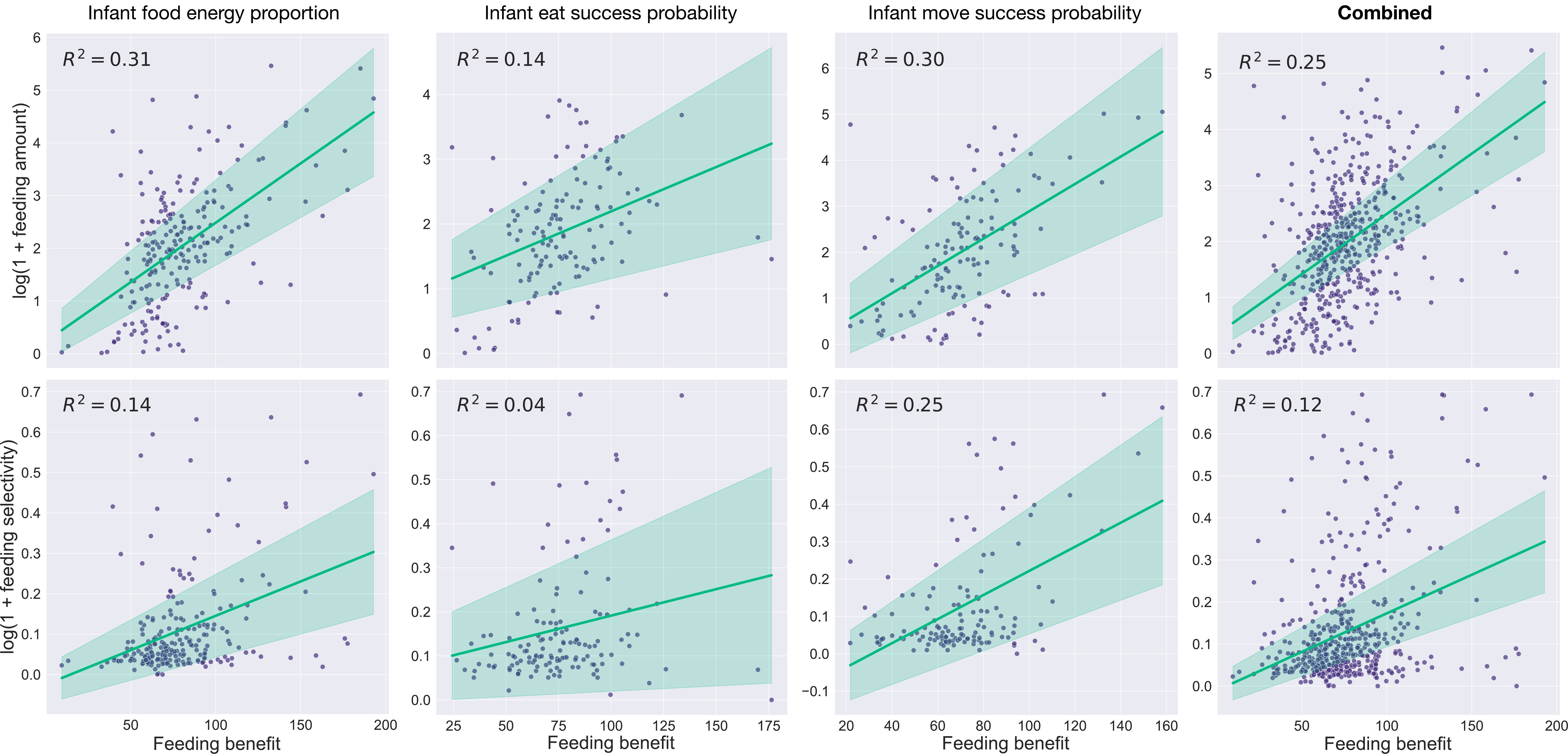}
    \caption{The relationship between the estimated benefit to infants of being fed and both the amount and selectivity of feeding observed, shown separately for each of the three experimental parameters we varied (and combined in the rightmost column). Each scatterplot point represents a single 500k-timestep simulation run (with values averaged over the final 50k timesteps); regression lines (with 95\% confidence intervals) are shown in green. Note that the $y$-axis shows $\log$(measure) for both amount and selectivity.}
    \label{fig:benefit-regressions}
\end{figure}

Having established an approximate measure of the benefit to reproductive fitness of being fed, we now test the relationship between this benefit and the actual feeding behaviour observed. If feeding behaviour is being evolved via kin selection, then we should expect two effects as we increase the importance of being fed to young agents' reproductive fitness:
\begin{itemize}
    \item The amount of feeding behaviour observed in the population should increase,
    \item The selectivity of feeding behaviour towards agents' own offspring should increase.
\end{itemize}
We measure the amount of feeding as simply the number of times the `feed' action was successfully used over the final 50k timesteps of each simulation run, divided by the average population size over the same period. To measure selectivity, we first take the proportion of these feeding events that were directed from parent to offspring. But this proportion is not necessarily a reliable measure of selectivity--it could be higher, for example, if infants stay closer to their parents on account of being less able to move. To account for this, we track all instances where one agent (agent A) is directly facing another agent (agent B), and compute the proportion of times where B is A's offspring. We can treat this proportion as a baseline for the relative level of feeding towards offspring that we would expect if there were no `real' selectivity; i.e. if agents were equally likely to use the feed action when facing any other agent. Our final measure of selectivity is then obtained by subtracting this baseline from the original proportion.
\begin{equation}\label{eq:selectivity}
    \text{selectivity} = \frac{\# \ \text{times offspring fed}}{\# \ \text{all feeding events}} - \frac{\# \ \text{times agent facing offspring}}{\# \ \text{times agent facing any agent}}
\end{equation}

\begin{table}[]
\centering
\begin{tabular}{@{}lllllll@{}}
\toprule
                                & \multicolumn{3}{l}{log(amount) $\sim$ benefit} & \multicolumn{3}{l}{log(selectivity) $\sim$ benefit} \\ \midrule
                                & $\beta$ & $r^2$ & $p$-val          & $\beta$   & $r^2$   & $p$-val           \\
varying infant food energy proportion   & 0.0225  & 0.312 & \textless{}0.001 & 0.00170   & 0.138   & \textless{}0.001  \\
varying infant eat success probability  & 0.0137  & 0.138 & \textless{}0.001 & 0.00120   & 0.043   & 0.014             \\
varying infant move success probability & 0.0297  & 0.297 & \textless{}0.001 & 0.00320   & 0.247   & \textless{}0.001  \\
\textbf{combined}                        & \textbf{0.0215}  & \textbf{0.254} & \textbf{\textless{}0.001} & \textbf{0.00180}   & \textbf{0.117}   & \textbf{\textless{}0.001}  \\ \bottomrule
\end{tabular}
\vspace{1ex}
\caption{Results of OLS linear regression analysis for feeding amount and selectivity against feeding benefit (to infants)}
\label{tab:benefit-regressions}
\end{table}

To test for the two trends listed above, we carried out a linear regression analysis (by the OLS method). We found significant positive correlations between feeding benefit and both $\log(\text{feeding amount})$ and $\log(\text{feeding selectivity})$, indicating that the two trends are indeed observed, and have an exponential shape. The regression results are given in Table \ref{tab:benefit-regressions}, and visualised in Figure \ref{fig:benefit-regressions}.


\subsection{The amount of feeding observed increases with the selectivity of feeding towards offspring}
\begin{figure}[t]
    \centering
    \includegraphics[width=\linewidth]{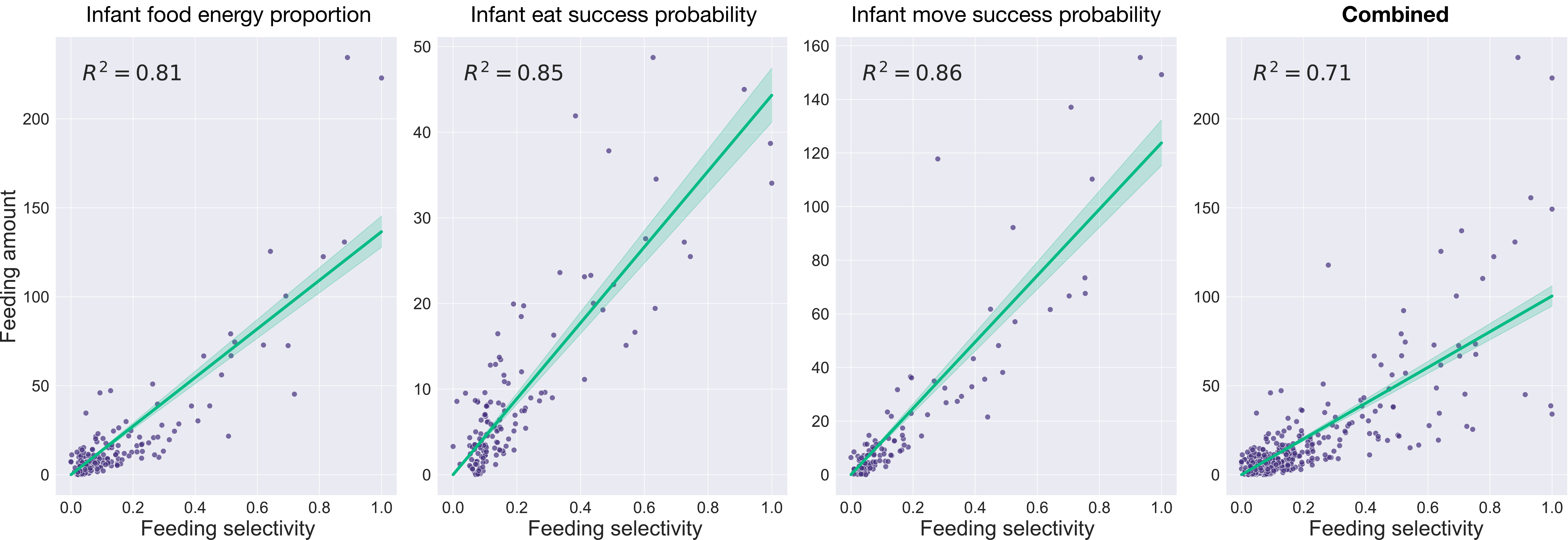}
    \caption{The relationship between the amount of feeding behaviour observed and its selectivity towards agents' own offspring, shown separately for each of the three experimental parameters we varied (and combined in the rightmost column). Each scatterplot point represents a single 500k-timestep simulation run (with values averaged over the final 50k timesteps); regression lines (with 95\% confidence intervals) are shown in green.}
    \label{fig:selectivity-regressions}
\end{figure}

\begin{table}[]
\centering
    \begin{tabular}{@{}llll@{}}
        \toprule
        & \multicolumn{3}{l}{amount $\sim$ selectivity} \\ \midrule
        & $\beta$ & $r^2$ & $p$-val \\
        varying infant food energy proportion   & 136.5  & 0.806 & \textless{}0.001 \\
        varying infant eat success probability  & 44.31  & 0.848 & \textless{}0.001 \\
        varying infant move success probability & 123.8  & 0.862 & \textless{}0.001 \\
        \textbf{combined} & \textbf{100.4}  & \textbf{0.708} & \textbf{\textless{}0.001} \\ \bottomrule
    \end{tabular}
    \vspace{1ex}
    \caption{Results of OLS linear regression analysis for feeding amount against feeding selectivity}
    \label{tab:selectivity-regressions}
\end{table}

If a behaviour evolves through kin selection, then we should expect to see a positive correlation between the overall prevalence of the behaviour and the extent to which it is used selectively towards agents' own kin. As an additional check, we therefore test the relationship between the amount and selectivity of feeding observed in our simulation runs. As is shown in Table \ref{tab:selectivity-regressions} and Figure \ref{fig:selectivity-regressions}, we find a very strong positive correlation between the two measures, across all three experimental parameters. 


\subsection{Kin selection operates through a combination of kin recognition and population viscosity}
\begin{figure}[t]
    \centering
    \includegraphics[width=\linewidth]{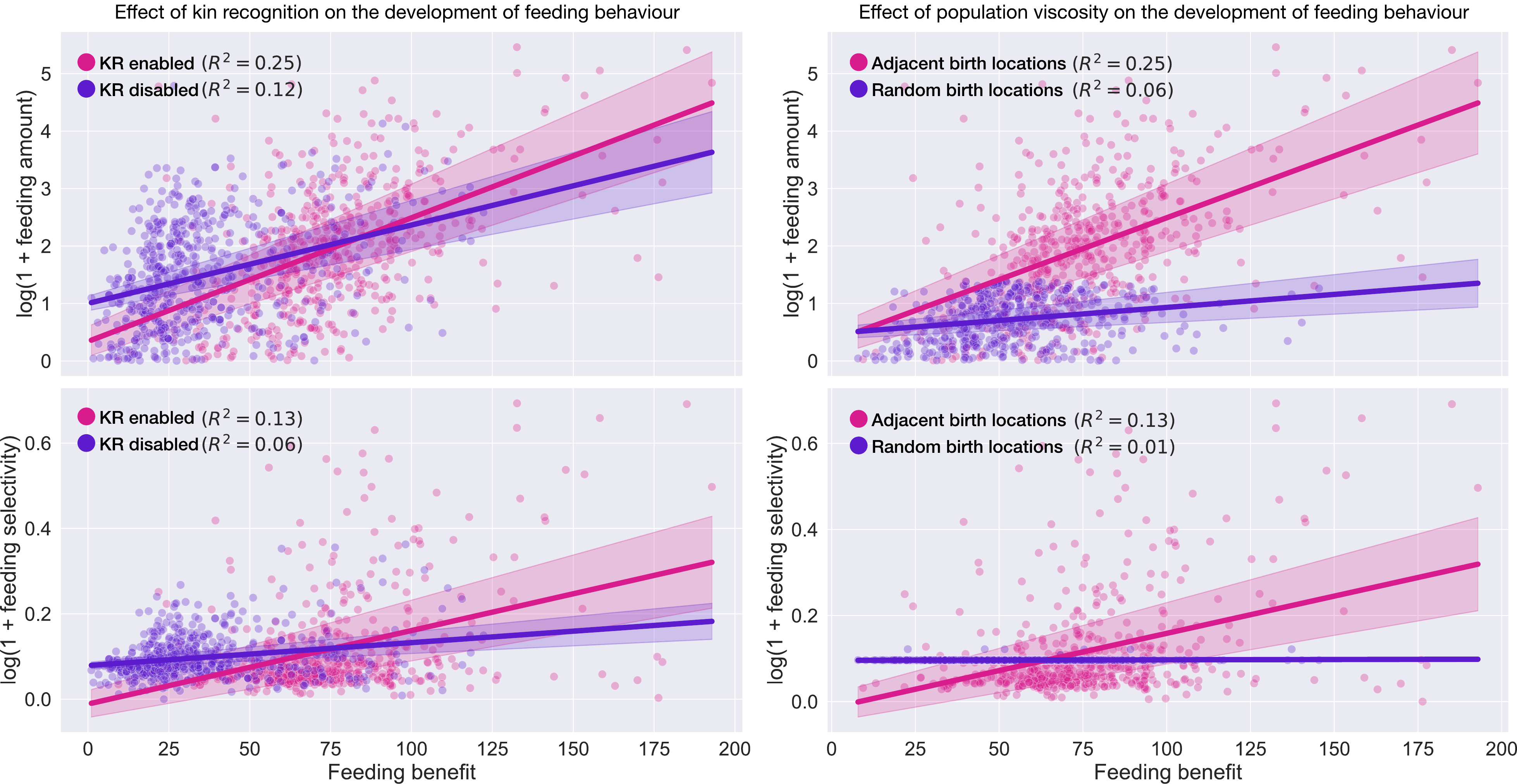}
    \caption{The effect of disabling kin recognition (left) and reducing population viscosity (right) on the relationships between feeding benefit and feeding amount and selectivity. Each dot represents a single 500k-timestep simulation run. Simulations were run over the same values of the three experimental parameters as before, and as before we ran 20 seeds per parameter value. 95\% confidence intervals are shown for all regression lines.}
    \label{fig:krpv}
\end{figure}

As mentioned briefly in Section \ref{sec:intro}, kin selection is understood to operate via two possible mechanisms: kin recognition, and population viscosity \cite{hamilton_genetical_1964_1,hamilton_genetical_1964_2}. In the case of kin recognition, individuals have some faculty for recognising their own kin, which allows them to target certain behaviours specifically towards them. In the absence of such a faculty, kin selection may still operate if the average movement of individuals from their birthplace is sufficiently slow. In such populations, the local social environment of an individual likely consists in large part of their relatives, simply by default--and so mere targeting of behaviour towards neighbours is sufficient for kin selection to occur. 

As a final experiment, we seek to understand which of these two mechanisms is responsible for the kin selection we observe in our virtual agent populations. To do this, we perform two independent ablations to our existing setup, and re-run the full set of simulations with each. To test the importance of kin recognition, we simply disable the observation feature that tells an agent whether each other agent they can see is one of their own offspring (see Figure \ref{fig:env-agent-diagram}; we retain the `is infant' feature). Population viscosity is a little trickier to intervene on directly. Our approach was to modify the reproduction dynamics. Previously, each new offspring was born into the tile their parent was facing at the time of reproduction (meaning parent and child were initially directly adjacent to one another). Instead, we now randomly select the birth location of all new offspring agents, meaning that a newborn is no more likely to be near to its parent as to any other agent in the population. 

Figure \ref{fig:krpv} shows the same regression analysis as performed previously for the two independent ablation runs. We can see that for both ablations, the relationships between feeding benefit ($b$) and the amount and selectivity of feeding are weakened. Interestingly, the effect is considerably stronger for randomised birth locations than for disabled kin recognition--suggesting that while both mechanisms play a role in the kin selection we observe, population viscosity is more important in facilitating the evolution of kin-selective feeding behaviour.

\section{Discussion}
In this paper, we have presented what is to our knowledge the first evidence of kin selection emerging from interactions between simulated agents undergoing neuroevolution in a simple but ecologically plausible environment. Focusing specifically on the phenomenon of agents feeding (donating energy to) their own infant offspring, we showed that this feeding behaviour saw greater adoption as a function of the fitness differential $rb - c$, as predicted by Hamilton's rule. We also found a very strong relationship between the prevalence of feeding behaviour observed and its selectivity towards agents' own offspring, supporting the idea that the use of the `feed' action emerged principally through a mechanism that reinforced its use towards kin but not non-kin. Finally, we investigated the mechanisms by which this kin selection process occurred, and found that while both kin recognition and population viscosity played a role, the latter had a more significant impact. 

As well as serving as a methodological demonstration for how kin selection and inclusive fitness can be studied using neuroevolution of simulated agents, we believe that our results may have some broader theoretical implications. The commonly accepted account of kin selection is made in terms of fitness maximisation from the perspective of individual genes shared by genetically related organisms. However, our simulations involve neither a proper notion of `gene' (or any discrete genetic subunit), nor an explicit notion of fitness maximisation. Agents are governed by, and pass to their offspring, a set of continuous neural network weights with no particular structure; agents survive and reproduce as long as they can maintain their internal energy above a certain level. The fact that we were still able to observe a kin selection phenomenon emerging suggests the possibility of a more general framing of the theory for such systems. One possible interpretation of our results is that by increasing the difficulty for infants to survive without being fed (i.e. increasing $rb - c$), we increased the extent to which the feeding of offspring was necessary for an agent to propagate a lineage over several generations. That is, as $rb - c$ increased, the probability increased for the genome of any agent \emph{not} engaging in offspring-feeding to disappear from the population, increasing the overall prevalence of the behaviour in the phenotype of the population.

While we do believe our results to be of some interest, we close by noting a number of limitations that could be addressed in future work that builds upon our framework. First, our agents' behaviour is governed by a very simple reactive architecture, in which actions are selected based purely on the current state observation, and there is no within-lifetime adaptation. This limits the complexity of the behavioural policies that agents can implement, and it would be interesting to explore a variant using (for example) recurrent architectures. For instance, an agent with memory could in principle `remember' which other agents are its own offspring without needing access to an explicit marker; or could adapt their behaviour to focus more on caregiving in later life. Secondly, our kin recognition mechanism is also very simplistic, only allowing agents to distinguish their own direct offspring. For a more complete test of the quantitative predictions made by Hamilton's rule, we would need to provide a continuous measure of the genetic relatedness between agents, in terms of their distance in neural parameter space. Finally, while not specifically a limitation of the current work, we believe that another fruitful research direction could be in exploring the evolution of more complex social behaviours within the same non-episodic neuroevolution framework. For example, could environment dynamics that change over time (instead of being stationary) foster the emergence of teaching or demonstration by parents for the benefit of their newborn offspring? 

\appendix
\section{Appendix}
All code is available at \href{https://anonymous.4open.science/r/EcoJAX-F6D4/}{this anonymised Github repo}.


\vspace{1em}

\textbf{Acknowledments: } This work was supported by the United Kingdom Research and Innovation (grant EP/S023208/1), EPSRC Centre for Doctoral Training in Robotics and Autonomous Systems (RAS). This work was also partially funded by the French National Research Agency (\url{https://anr.fr/}, project ECOCURL, Grant ANR-20-CE23-0006). This work also benefited from access to the HPC resources of IDRIS under the allocation 2023-[A0151011996] made by GENCI, using the Jean Zay supercomputer.

\bibliographystyle{splncs04}
\bibliography{references} 

\end{document}